# Antitopological magnetic textures in an antiferromagnetically coupled bilayer with frustration

Lewei Zhou, Jun Chen, Zhong Shen, Shuai Dong, and Xiaoyan Yao[*]

*Key Laboratory of Quantum Materials and Devices of Ministry of Education,*

*School of Physics, Southeast University, Nanjing 211189, China*

[*]Email: yaoxiaoyan@seu.edu.cn

**Abstract**

The bilayer skyrmion composed of upper and lower tightly coupled skyrmions on two layers with completely compensated topological charges (called as anti-topology here), has become one feasible improvement of conventional skyrmion to realize straight motion without skyrmion Hall effect, which has aroused great interest in practical applications. The present work investigates a general model (without external magnetic field) for the frustration-induced anti-topological bilayer magnetic textures with rich morphologies, and discusses the modulations of key parameters on the energy barrier and the current-driven dynamics. It is revealed that the interlayer coupling plays a key role in preventing distortion, and thus helps to reach a faster velocity. This model can be realized in various frustrated magnetic materials with antiferromagnetically coupled bilayer, providing a helpful guidance for the material design and application of topological magnetic textures.

Since the experimental discovery in 2009 [1], magnetic skyrmions have attracted significant attention as potential information carriers for next-generation spintronic devices [2-5], owing to their nanoscale size, topological stability and low threshold of driven current [6,7]. In most cases, magnetic skyrmions are induced by Dzyaloshinskii-Moriya interaction (DMI) in the materials lacking inversion symmetry. Meanwhile, skyrmions can also be stabilized in centrosymmetric materials by different factors, such as the dipolar interactions [8], the magnetic frustration [9] and the interaction between itinerant electron spins and localized spins [10]. In particular, the frustration-driven skyrmions may possess the full internal degrees of freedom, including different helicity and vorticity, and thus exhibit diverse morphologies [11], which are always absent in the DMI-induced cases. Besides the conventional skyrmions with the topological charge $Q_t = 1$, those with higher $Q_t$ may also exist in the frustrated systems [12-14].

To advance from basic research to practical applications, critical technical obstacles are still waiting to be overcome for skyrmions. Particularly, the skyrmion Hall effect stands out as one core issue hindering their device implementation, that is, the Magnus force causes skyrmions to deviate from the direction of driving current and subsequently annihilate upon drifting to the boundaries [15-17]. To eliminate the skyrmion Hall effect, much effort has been made, such as antiferromagnetic (AFM) skyrmion [18-21], skyrmion or antiskyrmion with unique helicity [22-24]. Among these, one more efficient and practical way is to construct a bilayer skyrmion composed of two tightly coupled textures with opposite topological charges (which are exactly cancelled, and thus can be called anti-topology [25]) in two layers, and thus leads to a stable linear transmission along the direction of driving current. The antiferromagnetically coupled ferromagnetic (FM) bilayer materials provide a superior playground to realize such anti-topological bilayer (ATB) skyrmions [26,27]. This scenario had been confirmed in experiments, and the fast current-induced straight motion at a speed up to 900 meters per second had been reported [28,29].

It is noteworthy that the ATB skyrmions in previous investigations were mainly induced by DMI, while the frustration-induced ATB textures have been rarely reported. However, the straight motion along the in-plane current direction and the better mobility than monolayer had been confirmed for the bilayer skyrmion in the previous dynamics investigation on the frustrated model at a specific parameter point [30]. In addition, it was predicted that bilayer-bimeron may exist in LaBr$_2$ bilayer without skyrmion Hall effect [31]. The above works only focus on specific materials at certain parameter points, and thus the systematical investigation on ATB magnetic textures in bilayer frustrated system is still absent.

In this work, we systematically investigate the ATB textures in the frustrated bilayer magnetic system with wide-range parameters by atomistic simulations and first-principles calculations. It is demonstrated that diverse ATB textures induced by frustration may coexist as the isolated metastable states, and the existent conditions are discussed in detail. The simulation on the dynamics driven by in-plane current verifies the straight motion without the skyrmion Hall effect. The modulation of key parameters on the current-driven velocity demonstrates similar trends to that of energy barrier. It is revealed that the interlayer coupling, which is absent in monolayer system, plays a key role in suppressing distortion and thus reaching a much faster velocity. Thereby, a general model for the frustration-induced ATB magnetic textures is provided, which can be realized in various frustrated materials with antiferromagnetically coupled bilayer, facilitating both fundamental physics and practical applications.

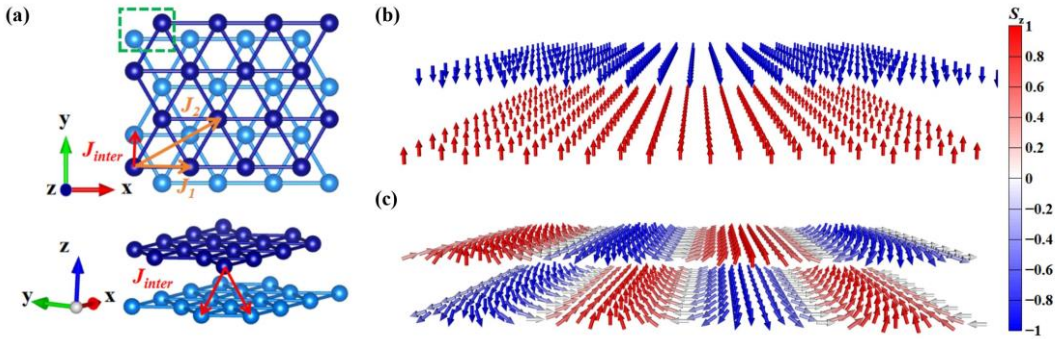

FIG.1. (a) Top and side views of the lattice structure. The dark/light blue balls represent the magnetic atoms in the top/bottom layer, and the green frame denotes the unit cell. Spin configurations for (b) A-AFM-1 and (c) Spiral-1 states.

Considering that the two-dimensional (2D) triangular lattice is the most typical structure with magnetic frustration [32], we focus on the bilayer magnetic system consisting of two triangular lattices. As usually observed in experiments, two triangular lattices tend to stack in the AB-mode [33,34], where the lattice sites of one triangular layer project to the centers of the other triangular plaquettes, and the inversion center locates at the bond center between the layers, as illustrated in Fig. 1(a). This structure was also confirmed to be energy-favored stacking sequence in theoretical works [35].

The Hamiltonian of this bilayer magnetic model is given by

$$H_{spin} = -J_1 \sum_{<i,j>} \mathbf{S}_i \cdot \mathbf{S}_j - J_2 \sum_{<i,k>} \mathbf{S}_i \cdot \mathbf{S}_k - J_{inter} \sum_{<i,m>} \mathbf{S}_i \cdot \mathbf{S}_m - K \sum_i (S_i^z)^2 \qquad (1)$$

Here, $\mathbf{S}_i$ is the normalized classical spin at the $i$th site. $J_1$ and $J_2$ describe the exchange couplings between the intralayer first and second nearest-neighboring spins, while $J_{inter}$ denotes the exchange coupling between the interlayer nearest-neighboring spins, as marked in Fig. 1(a). Here $J_1 = 1$ is fixed as the energy unit, and all the parameters are simplified with reduced units. $K$ is the strength of the single-ion anisotropy of easy-axis ($K > 0$) or easy-plane ($K < 0$) type. Since centrosymmetric bilayer magnets can only possess the layer-dependent staggered DMI, which is almost compensated for the whole system. It was reported that this layer-dependent staggered DMI only takes obvious effect with weak $J_{inter}$, and stabilizes the bilayer skyrmion crystal with non-canceling topological charges under an external magnetic field [36]. Therefore, DMI is not considered here. Based on this model [Eq. (1)], we perform atomistic simulations with Landau-Lifshitz-Gilbert (LLG) equation, as implemented in the SPIRIT package [37], to explore the spin textures and the spin dynamics (see Note 1 in the Supplemental Material (SM) [38] for details).

In Eq. (1), the FM $J_1 > 0$ tends to produce an intralayer FM state, while $J_{inter} < 0$ induces an interlayer AFM order. Thus, an A-AFM state with antiferromagnetically coupled FM layers is generated, including A-AFM-1 (all the spins normal to the plane) for $K > 0$ and A-AFM-2 (all the spins within the plane) for $K < 0$, as illustrated in Fig. 1(b) and Fig. S1(a) in SM [38]. The AFM $J_2 < 0$ introduces intralayer frustration, which leads to the intralayer ground state transition from FM to spiral state, including Spiral-1 (spiral plane normal to $xy$ plane) for $K > 0$ and Spiral-2 (spiral plane within $xy$ plane) for $K < 0$, as illustrated in Fig. 1(c) and Fig. S1(b) in SM [38]. Meanwhile, the frustration provides necessary conditions to stabilize isolated ATB magnetic textures. Since $J_{inter}$ plays the role of built-in magnetic field, the external magnetic field is unnecessary and not considered in the present work.

As exhibited in Fig. 2, for each ATB magnetic texture, the similar textures in the two layers with opposite topological charges and a helicity difference of π, are tightly coupled due to the interlayer AFM interaction. Owing to the total topological charge $Q_t = 0$ invariably, the topological charge of the top texture is represented by $Q$ especially. Due to frustration, these ATB textures possess rich internal degrees of freedom, and thus exhibit distinct morphologies. The skyrmions of Bloch, Néel types and intermediate helicity, antiskyrmion and even the textures with higher $|Q|$ may coexist.

For convenience, these textures existing at $K > 0$ are called skyrmionic textures, and those at $K < 0$ are called bimeronic textures. Owing to the frustration origin, the ATB skyrmionic textures with the same $|Q|$ are degenerate in energy, and so are ATB bimeronic textures. Due to the high internal degrees of freedom, the frustration-induced topological textures on two layers are easier to match. Such a diversity of coexisting ATB textures is hardly accessible in DMI-induced cases.

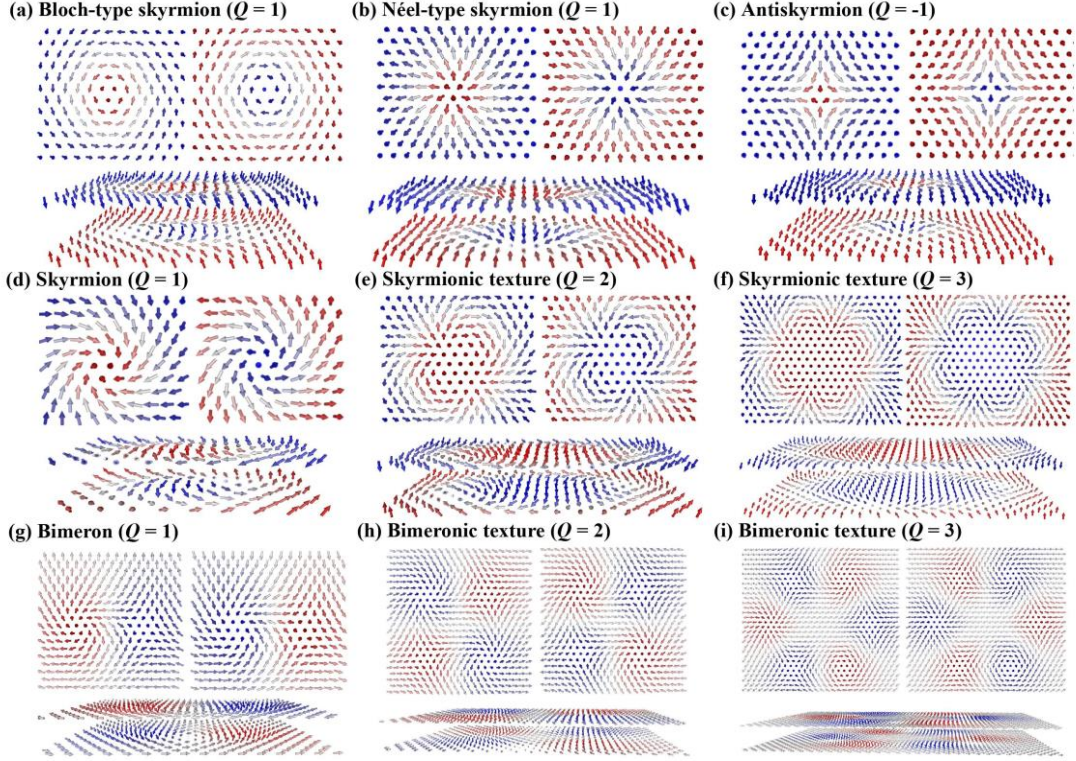

FIG. 2. The ATB textures obtained by simulations. (a) Bloch-type skyrmion ($Q = 1$), (b) Néel-type skyrmion ($Q = 1$) and (c) antiskyrmion ($Q = -1$). Skyrmionic textures of (d) $Q = 1$, (e) $Q = 2$ and (f) $Q = 3$ with an intermediate helicity. Bimeronic textures of (g) $Q = 1$, (h) $Q = 2$, and (i) $Q = 3$. The skyrmionic textures (a)-(f) are obtained with $J_2 = -0.4$, $J_{\text{inter}} = -2$ and $K = 0.01$. The bimeronic textures (g)-(i) are obtained with $J_2 = -0.4$, $J_{\text{inter}} = -2$ and $K = -0.01$. For each figure, the two top panels present the textures on upper and lower layers, which are shown together in the bottom panel. The same color bar is used as in Fig. 1.

All these ATB magnetic textures are metastable. Their existence can be confirmed by the energy barrier during the collapse [39,40], which is calculated by using the geodesic nudged elastic band (GNEB) method [37,41]. Figs. 3(a) and 3(b) plots the energy variation during the evolution procedure from the metastable ATB textures to the A-AFM ground state at a certain parameter point. It is seen that the texture with higher $|Q|$ possess higher energy, and thus there are three energy levels for $|Q| = 1, 2$ and 3. During the annihilation process with $|Q|$ decreasing by 1 in sequence down to 0, the energy goes down these levels one by one, as illustrated by the intermediate states in Fig. 3(c). At a certain parameter point, for each energy step from $|Q|$ to $|Q| - 1$, there is a uniform energy barrier ($h$) that needs to be overcome, and $h$ increases slightly for a larger $|Q|$, as displayed in Figs. 3(a) and 3(b).

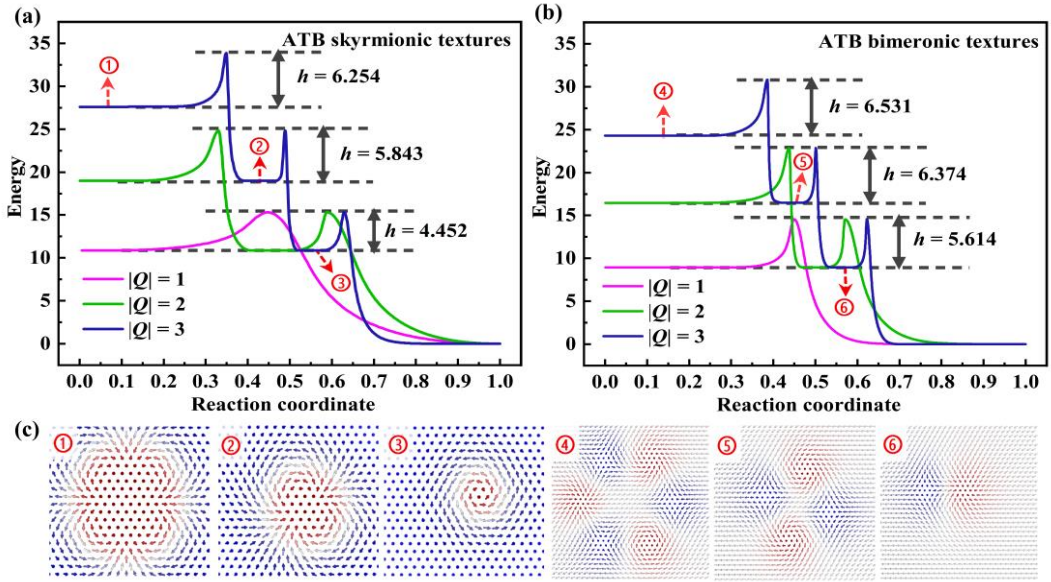

FIG.3. The energy variations during the collapse to the A-AFM ground state with $J_2 = -0.4$ and $J_{inter} = -2$ for (a) skyrmionic textures ($K = 0.01$) and (b) bimeronic textures ($K = -0.01$). (c) The intermediate images marked in (a) and (b), where only the spin configurations in the top layer are plotted and the same color bar is used as in Fig. 1.

Taking the ATB skyrmion and bimeron with $Q = 1$ as examples, the existent conditions and dynamics are discussed in detail. The phase diagram including ground states and metastable ATB textures is determined by the competition between the frustration ($J_2$), anisotropy ($K$) and interlayer $J_{inter}$. As demonstrated in Fig. 4(a). With increasing $J_2$, the ground state transforms from the A-AFM-1/A-AFM-2 state into the Spiral-1/Spiral-2 state for $K > 0/K < 0$. When $K > 0$, the range of A-AFM-1 state expands and the Spiral-1 state is suppressed as $K$ increases. When $K < 0$, all the spins of the A-AFM-2 and Spiral-2 states are restricted within $xy$ plane due to easy-plane anisotropy, but the spin variation within the $xy$ plane is not restricted due to the in-plane isotropy. Thus, $K$ has no effect on the phase boundary between the A-AFM-2 and Spiral-2 states.

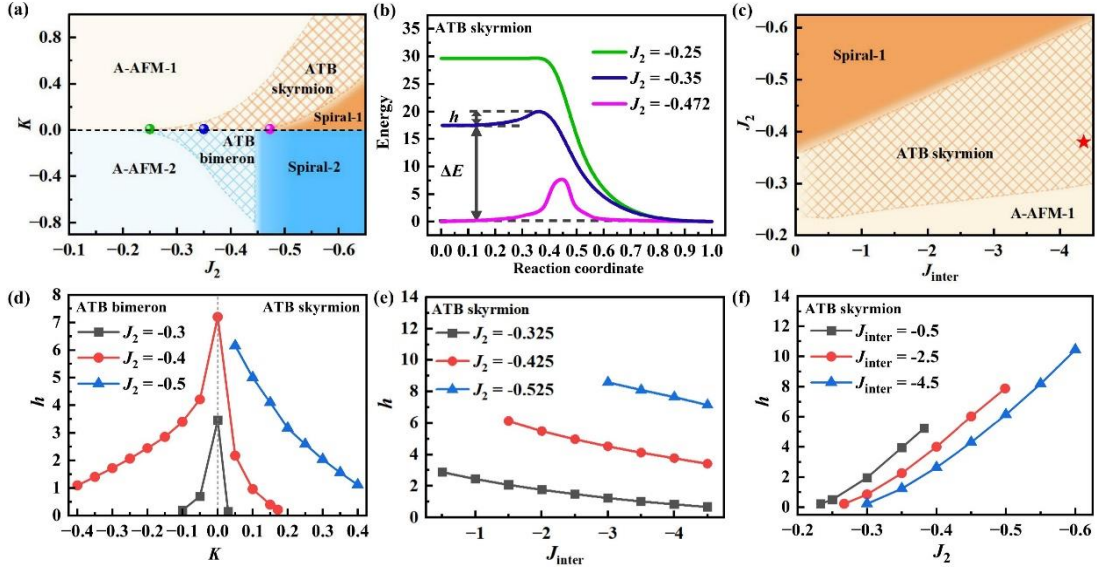

FIG. 4 Phase diagrams and energy barrier's modulation for the ATB skyrmion and bimeron of $Q = 1$. (a) Phase diagram with fixed $J_{inter} = -2$ in the $K$-$J_2$ parameter space. (b) The energy variations during the collapse from ATB skyrmion ($Q = 1$) to the A-AFM-1 state with $J_{inter} = -2$, $K = 0.01$ and different $J_2$. The corresponding parameter points are marked in (a). (c) Phase diagram with fixed $K = 0.01$ in the $J_2$-$J_{inter}$ parameter space. Red star: the case for $Cr_2HfC_2H_2$. In (a) and (c), the ground states (determined by energy comparison) are distinguished by background color, while the metastable regions of ATB textures are marked by grid. $h$ as a function of (d) $K$ ($J_{inter} = -2$), (e) $J_{inter}$ ($K = 0.01$) and (f) $J_2$ ($K = 0.01$).

Fig. 4(b) displays the energy evolution obtained by GNEB method from the ATB skyrmion to the A-AFM-1 ground state. Two characteristic quantities, i.e. $h$ and the energy difference between the ATB skyrmion and A-AFM-1 state ($\Delta E$), are noteworthy. At the boundary of the ATB skyrmion close to the A-AFM-1 state, it is observed that $\Delta E > 0$ but $h$ approaches 0, indicating spontaneous annihilation of any skyrmions. At the other boundary close to Spiral-1 state, $h > 0$ but $\Delta E$ approaches 0, which implies the transition of ground state. Thus, the metastable region of ATB skyrmion can be identified with $\Delta E > 0$ and $h > 0$, which just locates within the A-AFM-1 phase close to the boundary of the Spiral-1 state, as shown in Fig. 4(a). The metastable region of ATB bimeron is also located within the A-AFM-2 phase close to the boundary of the Spiral-2 state.

As presented in Fig. 4(a), when $K > 0$, the region of ATB skyrmion is continually pushed toward higher frustration by increasing $K$. However, as $K < 0$, an enhancement in $|K|$ suppresses the region of ATB bimeron because the internal spins of bimeron are not entirely aligned within the easy plane. $J_{inter}$, as the most important character of bilayer in differentiating from monolayer, provides a built-in magnetic field, which expands the A-AFM ground state, and thus effectively extends the existent range of ATB textures to higher frustration, as displayed in Fig. 4(c). In the existent region of ATB textures, the modulation of key parameters on $h$ is discussed in detail. Fig. 4(d) exhibits that $h$ decreases with $|K|$ increasing, so $|K| = 0.01$ is fixed in most discussions. Since $J_{inter}$ suppresses the noncollinear arrangement of spins, $h$ decreases gradually with $|J_{inter}|$ rising, as shown in Fig. 4(e). Fig. 4(f) demonstrates that strengthening $J_2$ enhances frustration and thus raises $h$ obviously. To avoid repetition, some figures of ATB bimeron are presented in Note 2 of SM [38].

One significant advantage of these ATB textures is the straight motion driven by spin-polarized electric current without skyrmion Hall effect, which is confirmed in the present simulation. When the current is injected along the $x$ direction in both the top and bottom layers, all the ATB textures move in a straight line along the $x$ direction, and the textures with the same $|Q|$ show identical velocity ($v$) at a certain parameter point. As plotted in Fig. 5(a), $v$ increases with $J_2$ and decreases with $K$ or $J_{inter}$ at a certain magnitude of spin transfer torque ($u = 0.1$ m/s), which is consistent with the variation of $h$ modulated by these parameters.

It is well-known that the distortion is one key obstacle in the current-driven motion of topological magnetic texture. $J_{inter}$ can enhance the stiffness of ATB texture and suppress its distortion efficiently. When the current is applied, the ATB skyrmion is driven to move, and simultaneously the texture is stretched along the direction normal to current. Here the ratio of height to width ($R$) is used to characterize the distortion. Fig 5(b) presents the time ($t$) evolution of $R$. At a certain $u$, if $J_{inter}$ is relatively weak, $R$ increases continuously until the texture can not maintain the particle-like feature any more. If $J_{inter}$ is relatively strong, $R$ converges to a stable value ($R_s$) quickly, and then the robust shape is kept during the subsequent uniform motion. As shown in Fig. 5(c), at a certain $J_{inter}$, $R_s$ grows slowly first and fast afterwards with $u$ increasing, and the turning point of growth rate can be estimated as the critical $u$ ($u_c$). It is obvious that $J_{inter}$ can suppress the distortion efficiently, and thus $u_c$ increases with $J_{inter}$ strengthening, as demonstrated in the inset of Fig. 5(d). $u_c = 0.2$ m/s with $J_{inter} = -0.5$, while $u_c = 2.7$ m/s with $J_{inter} = -4.5$, which means much larger current can be withstood via providing stronger $J_{inter}$. Fig. 5(d) exhibits the current-driven velocity of the ATB skyrmion with $R$s < 3. It is seen that $v$ depends on $u$ monotonously. Therefore, much higher $v$ can be obtained at larger $u$, which can be realized with stronger $J_{inter}$.

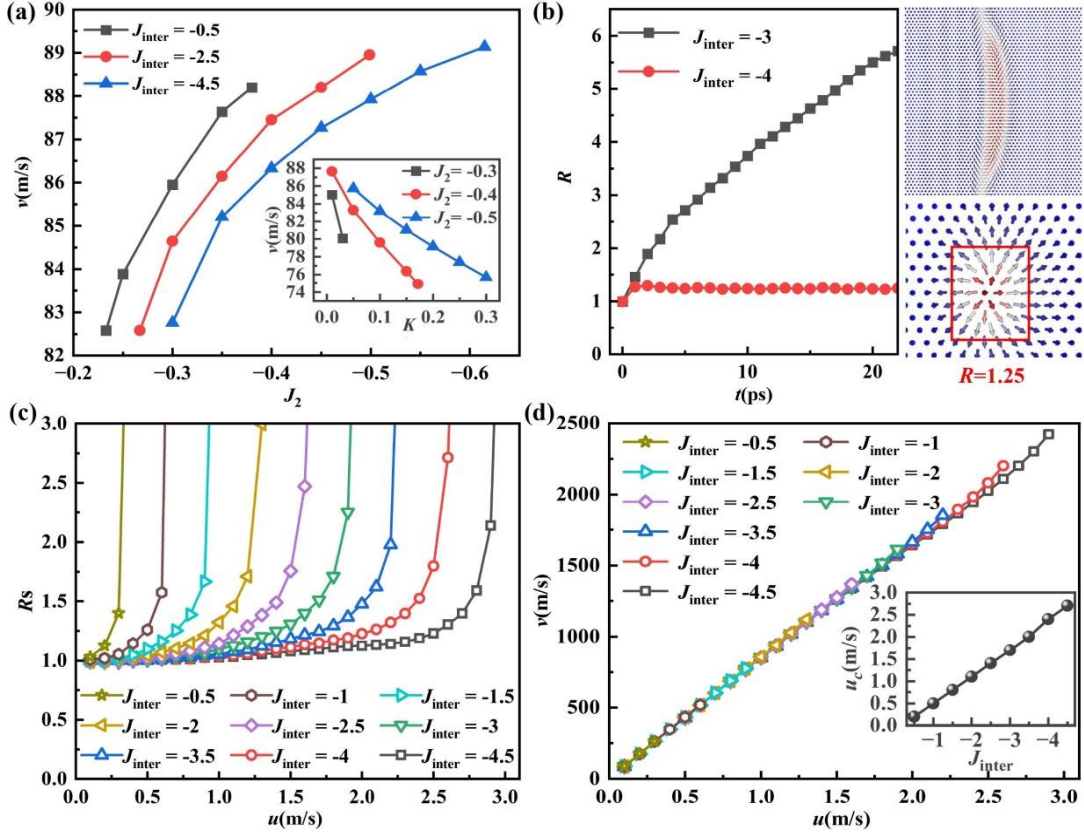

FIG.5. The current-driven dynamics of the isolated ATB skyrmion. (a) The dependence of velocity $v$ on $J_2$ with $K = 0.01$. Inset: $v$ as a function of $K$ with $J_{inter} = -2$. The magnitude of spin transfer torque $u = 0.1$ m/s. (b) The time ($t$) evolution of length-to-width ratio $R$ with $u = 2.05$ m/s. The right above/down inset presents the disperse/convergent state at $t = 22$ ps, where only the top texture is plotted and the same color bar is used as in Fig. 1. (c) $R$s and (d) $v$ as a function of $u$ for different $J_{inter}$ during the uniform motion. The inset in (d) presents $u_c$ as a function of $J_{inter}$. (b-d) $J_2 = -0.35$ and $K = 0.01$.

The behavior similar to the ATB skyrmion can be observed for ATB bimerons (see Note 2 in SM [38] for more details). Thereby, the above discussion presents a general scenario for ATB textures in the bilayer magnets. The case with small $J_{inter}$ could be realized in the 2D van der Waals (vdW) materials. For example, it was reported that bilayer LaBr$_2$ with $J_1$ = 8.651 meV, $J_2$ = -2.339 meV, $J_{inter}$ = -0.522 meV, and $K$ = -0.024 meV, i.e., $J_2$ = -0.270, $J_{inter}$ = -0.060, and $K$ = -0.003 in units of $J_1$, falls within the region of ATB bimeron, and the bilayer bimeron was observed with the skyrmion Hall effect suppressed [31].

On the other end, the case with strong $J_{inter}$ can be realized in the 2D ordered double transition metals carbides (MXenes) materials [42]. For instance, $Cr_2HfC_2H_2$ possesses a centrosymmetric magnetic bilayer structure, namely two Cr magnetic layers separated by a non-magnetic C-Hf-C layer. First-principles calculations show that under 5.5% in-plane biaxial compressive strain, $J_1$ = 8.600 meV, $J_2$ = -3.254 meV, $J_{inter}$ = -37.530 meV and $K$ = 0.085 meV, i.e., $J_2$ = -0.378, $J_{inter}$ = -4.364, $K$ = 0.010 in units of $J_1$, just falling within the region of ATB skyrmion, as marked in Fig. 4(c). The atomistic simulation based on these parameters confirms the coexistence of ATB skyrmionic textures with distinct $Q$ (see Note 3 in SM [38] for more details).

In conclusion, the frustration-induced ATB textures with rich morphologies may exist as the isolated metastable state in the bilayer magnetic system. Their energy barriers exhibit a positive correlation with frustration and a negative correlation with interlayer coupling and anisotropy. The simulation on the current-driven dynamics confirms the straight motion without the skyrmion Hall effect, and the velocity dependence on magnetic parameters follows the same trend as that of the energy barrier. It is revealed that the interlayer coupling may enhance the stiffness of ATB textures and suppress the distortion efficiently, and thus enable the texture to endure higher current and reach a much faster velocity. The systematical investigation in the present work provides a general model for the ATB magnetic textures in the frustrated bilayer. Both MXenes and vdW materials are proposed as candidate platforms for ATB textures. The cases with strong interlayer coupling could be realized in MXenes, such as $Cr_2HfC_2H_2$, while those with weak interlayer coupling could be realized in the 2D vdW materials.

**ACKNOWLEDGEMENTS**

We thank Yijun Yang of Southeast University for fruitful discussions. This work is supported by National Natural Science Foundation of China (Grant Nos. 12574090, 12504123, 124B2064), Jiangsu Funding Program for Excellent Postdoctoral Talent and the Postdoctoral Fellowship Program China of CPSF (Grant No. GZB20250766). We thank the Big Data Center of Southeast University for providing the facility support on the numerical calculations.

# REFERENCES

[1] S. Muehlbauer, B. Binz, F. Jonietz, C. Pfleiderer, A. Rosch, A. Neubauer, R. Georgii, and P. Boeni, Skyrmion lattice in a chiral magnet, Science **323**, 915 (2009).

[2] A. Fert, V. Cros, and J. Sampaio, Skyrmions on the track, Nat. Nanotechnol. **8**, 152 (2013).

[3] X. Zhang, M. Ezawa, and Y. Zhou, Magnetic skyrmion logic gates: conversion, duplication and merging of skyrmions, Sci. Rep. **5**, 9400 (2015).

[4] A. Fert, N. Reyren, and V. Cros, Magnetic skyrmions: advances in physics and potential applications, Nat. Rev. Mater. **2**, 17031 (2017).

[5] X. Zhang, Y. Zhou, K. M. Song, T.-E. Park, J. Xia, M. Ezawa, X. Liu, W. Zhao, G. Zhao, and S. Woo, Skyrmion-electronics: writing, deleting, reading and processing magnetic skyrmions toward spintronic applications, J. Phys.: Condens. Matter **32**, 143001 (2020).

[6] X. Z. Yu, N. Kanazawa, W. Z. Zhang, T. Nagai, T. Hara, K. Kimoto, Y. Matsui, Y. Onose, and Y. Tokura, Skyrmion flow near room temperature in an ultralow current density, Nat. Commun. **3**, 988 (2012).

[7] N. Nagaosa and Y. Tokura, Topological properties and dynamics of magnetic skyrmions, Nat. Nanotechnol. **8**, 899 (2013).

[8] X. Yu, M. Mostovoy, Y. Tokunaga, W. Zhang, K. Kimoto, Y. Matsui, Y. Kaneko, N. Nagaosa, and Y. Tokura, Magnetic stripes and skyrmions with helicity reversals, Proc. Natl. Acad. Sci. U. S. A. **109**, 8856 (2012).

[9] T. Okubo, S. Chung, and H. Kawamura, Multiple-q states and the skyrmion lattice of the triangular-lattice heisenberg antiferromagnet under magnetic fields, Phys. Rev. Lett. **108**, 017206 (2012).

[10] R. Ozawa, S. Hayami, and Y. Motome, Zero-field skyrmions with a high topological number in itinerant magnets, Phys. Rev. Lett. **118**, 147205 (2017).

[11] X. Yao and S. Dong, Vector vorticity of skyrmionic texture: an internal degree of freedom tunable by magnetic field, Phys. Rev. B **105**, 014444 (2022).

[12] H. L. Hu, Z. Shen, Z. Chen, X. P. Wu, T. T. Zhong, and C. S. Song, High-topological-number skyrmions with tunable diameters in two-dimensional frustrated $J_1$-$J_2$ magnets, Appl. Phys. Lett. **125**, 092402 (2024).

[13] A. O. Leonov and M. Mostovoy, Multiply periodic states and isolated skyrmions in an anisotropic frustrated magnet, Nat. Commun. **6**, 8275 (2015).

[14] X. C. Zhang, J. Xia, Y. Zhou, X. X. Liu, H. Zhang, and M. Ezawa, Skyrmion dynamics in a frustrated ferromagnetic film and current-induced helicity locking-unlocking transition, Nat. Commun. **8**, 1717 (2017).

[15] K. Litzius, I. Lemesh, B. Krueger, P. Bassirian, L. Caretta, K. Richter, F. Buettner, K. Sato, O. A. Tretiakov, J. Foerster, R. M. Reeve, M. Weigand, L. Bykova, H. Stoll, G. Schuetz, G. S. D. Beach, and M. Klaeui, Skyrmion Hall effect revealed by direct time-resolved x-ray microscopy, Nat. Phys. **13**, 170 (2017).

[16] W. J. Jiang, X. C. Zhang, G. Q. Yu, W. Zhang, X. Wang, M. B. Jungfleisch, J. E. Pearson, X. M. Cheng, O. Heinonen, K. L. Wang, Y. Zhou, A. Hoffmann, and S. G. E. te Velthuis, Direct observation of the skyrmion Hall effect, Nat. Phys. **13**, 162 (2017).

[17] S. Woo, K. M. Song, X. Zhang, Y. Zhou, M. Ezawa, X. Liu, S. Finizio, J. Raabe, N. J. Lee, S.-I. Kim, S.-Y. Park, Y. Kim, J.-Y. Kim, D. Lee, O. Lee, J. W. Choi, B.-C. Min, H. C. Koo, and J. Chang,

Current-driven dynamics and inhibition of the skyrmion Hall effect of ferrimagnetic skyrmions in GdFeCo films, Nat. Commun. **9**, 959 (2018).

[18] J. Barker and O. A. Tretiakov, Static and dynamical properties of antiferromagnetic skyrmions in the presence of applied current and temperature, Phys. Rev. Lett. **116**, 147203 (2016).

[19] C. D. Jin, C. K. Song, J. B. Wang, and Q. F. Liu, Dynamics of antiferromagnetic skyrmion driven by the spin Hall effect, Appl. Phys. Lett. **109**, 182404 (2016).

[20] X. Zhang, Y. Zhou, and M. Ezawa, Antiferromagnetic skyrmion: stability, creation and manipulation, Sci. Rep. **6**, 24795 (2016).

[21] L. C. Shen, J. Xia, G. P. Zhao, X. C. Zhang, M. Ezawa, O. A. Tretiakov, X. X. Liu, and Y. Zhou, Dynamics of the antiferromagnetic skyrmion induced by a magnetic anisotropy gradient, Phys. Rev. B **98**, 134448 (2018).

[22] B. Q. Dai, D. Wu, S. A. Razavi, S. J. Xu, H. R. He, Q. Y. Shu, M. Jackson, F. Mahfouzi, H. S. Huang, Q. J. Pan, Y. Cheng, T. Qu, T. Y. Wang, L. X. Tai, K. Wong, N. Kioussis, and K. L. Wang, Electric field manipulation of spin chirality and skyrmion dynamics, Sci. Adv. **9**, eade6836 (2023).

[23] Z. L. He, K. Y. Dou, W. H. Du, Y. Dai, B. B. Huang, and Y. D. Ma, Mixed Bloch-Néel type skyrmions in a two-dimensional lattice, Phys. Rev. B **109**, 024420 (2024).

[24] S. Y. Huang, C. Zhou, G. Chen, H. Y. Shen, A. K. Schmid, K. Liu, and Y. Z. Wu, Stabilization and current-induced motion of antiskyrmion in the presence of anisotropic Dzyaloshinskii-Moriya interaction, Phys. Rev. B **96**, 144412 (2017).

[25] A. P. Reddy, D. N. Sheng, A. Abouelkomsan, E. J. Bergholtz, and L. Fu, Anti-topological crystal and non-Abelian liquid in twisted semiconductor bilayers, arXiv:2411.19898 (2024).

[26] X. Zhang, Y. Zhou, and M. Ezawa, Magnetic bilayer-skyrmions without skyrmion Hall effect, Nat. Commun. **7**, 10293 (2016).

[27] Q. R. Cui, Y. M. Zhu, J. H. Liang, P. Cui, and H. X. Yang, Antiferromagnetic topological magnetism in synthetic van der Waals antiferromagnets, Phys. Rev. B **107**, 064422 (2023).

[28] R. Y. Chen, Q. R. Cui, L. Han, X. T. Xue, J. H. Liang, H. Bai, Y. J. Zhou, F. Pan, H. X. Yang, and C. Song, Controllable generation of antiferromagnetic skyrmions in synthetic antiferromagnets with thermal effect, Adv. Funct. Mater. **32**, 2111906 (2022).

[29] V. Pham, N. Sisodia, I. Di Manici, J. Urrestarazu-Larrañaga, K. Bairagi, J. Pelloux-Prayer, R. Guedas, L. D. Buda-Prejbeanu, S. Auffret, A. Locatelli, T. O. Mentes, S. Pizzini, P. Kumar, A. Finco, V. Jacques, G. Gaudin, and O. Boulle, Fast current-induced skyrmion motion in synthetic antiferromagnets, Science **384**, 307 (2024).

[30] J. Xia, X. C. Zhang, M. Ezawa, Z. P. Hou, W. H. Wang, X. X. Liu, and Y. Zhou, Current-Driven dynamics of frustrated skyrmions in a synthetic antiferromagnetic bilayer, Phys. Rev. Appl. **11**, 044046 (2019).

[31] W. Sun, W. X. Wang, H. Li, X. N. Li, Z. Y. Yu, Y. Bai, G. B. Zhang, and Z. X. Cheng, LaBr2 bilayer multiferroic moire superlattice with robust magnetoelectric coupling and magnetic bimerons, npj Comput. Mater. **8**, 159 (2022).

[32] J. Khatua, B. Sana, A. Zorko, M. Gomilsek, K. Sethupathi, M. S. R. Rao, M. Baenitz, B. Schmidt, and P. Khuntia, Experimental signatures of quantum and topological states in frustrated magnetism, Phys. Rep. **1041**, 1 (2023).

[33] H. Ge, T. Li, S. E. Nikitin, N. Zhao, F. Li, H. Bu, J. Yuan, J. Chen, Y. Fu, J. Yang, L. Wang, P. Miao, Q. Zhang, I. Puente-Orench, A. Podlesnyak, J. Sheng, and L. Wu, Magnetic structure and Ising-like

antiferromagnetism in the bilayer triangular lattice compound NdZnPO, Phys. Rev. B **110**, 054443 (2024).

[34] S. Nakatsuji, H. Tonomura, K. Onuma, Y. Nambu, O. Sakai, Y. Maeno, R. T. Macaluso, and J. Y. Chan, Spin disorder and order in quasi-2D triangular heisenberg antiferromagnets: comparative study of $FeGa_2S_4$, $Fe_2Ga_2S_5$, and $NiGa_2S_4$, Phys. Rev. Lett. **99**, 157203 (2007).

[35] Q. Q. Li, W. W. Liu, Z. K. Ding, H. Pan, X. H. Cao, W. H. Xiao, N. N. Luo, J. Zeng, L. M. Tang, B. Li, K. Q. Chen, and X. D. Duan, Stacking- and strain-dependent magnetism in Janus CrSTe bilayer, Appl. Phys. Lett. **122**, 121902 (2023).

[36] S. Hayami, Skyrmion crystal and spiral phases in centrosymmetric bilayer magnets with staggered Dzyaloshinskii-Moriya interaction, Phys. Rev. B **105**, 014408 (2022).

[37] G. P. Mueller, M. Hoffmann, C. Disselkamp, D. Schuerhoff, S. Mavros, M. Sallermann, N. S. Kiselev, H. Jonsson, and S. Bluegel, Spirit: multifunctional framework for atomistic spin simulations, Phys. Rev. B **99**, 224414 (2019).

[38] See Supplementary Material for the simulation details; the energy barrier and dynamics of ATB bimerons; the calculations on $Cr_2HfC_2H_2$; the dynamics of ATB skyrmionic textures with high $Q$; and the potential application and modulation, which cites Refs. [11,22,23,37,43-52].

[39] B. Heil, A. Rosch, and J. Masell, Universality of annihilation barriers of large magnetic skyrmions in chiral and frustrated magnets, Phys. Rev. B **100**, 134424 (2019).

[40] A. Aldarawsheh, I. L. Fernandes, S. Brinker, M. Sallermann, M. Abusaa, S. Blügel, and S. Lounis, Emergence of zero-field non-synthetic single and interchained antiferromagnetic skyrmions in thin films, Nat. Commun. **13**, 7369 (2022).

[41] P. F. Bessarab, V. M. Uzdin, and H. Jónsson, Method for finding mechanism and activation energy of magnetic transitions, applied to skyrmion and antivortex annihilation, Comput. Phys. Commun. **196**, 335 (2015).

[42] B. Anasori, Y. Xie, M. Beidaghi, J. Lu, B. C. Hosler, L. Hultman, P. R. C. Kent, Y. Gogotsi, and M. W. Barsoum, Two-dimensional, ordered, double transition metals carbides (MXenes), ACS Nano **9**, 9507 (2015).

[43] J. H. Guo, Y. Hou, J. Xia, X. Zhang, P. W. T. Pong, and Y. Zhou, Dynamic properties of a ferromagnetic skyrmion in an in-plane magnetic field, J. Appl. Phys. **131**, 073901 (2022).

[44] J. C. Martinez, W. S. Lew, W. L. Gan, and M. B. A. Jalil, Theory of current-induced skyrmion dynamics close to a boundary, J. Magn. Magn. Mater. **465**, 685 (2018).

[45] G. Kresse and J. Furthmüller, Efficient iterative schemes for ab initio total-energy calculations using a plane-wave basis set, Phys. Rev. B **54**, 11169 (1996).

[46] S. Bae, Y. G. Kang, M. Khazaei, K. Ohno, Y. H. Kim, M. J. Han, K. J. Chang, and H. Raebiger, Electronic and magnetic properties of carbide MXenes-the role of electron correlations, Mater. Today Adv. **9**, 100118 (2021).

[47] Y. G. Zhang, Z. Cui, B. S. Sa, N. H. Miao, J. Zhou, and Z. M. Sun, Computational design of double transition metal MXenes with intrinsic magnetic properties, Nanoscale Horiz. **7**, 276 (2022).

[48] J. Chen, J. J. Liang, J. H. Yu, M. H. Qin, Z. Fan, M. Zeng, X. B. Lu, X. S. Gao, S. Dong, and J-M Liu, Dynamics of distorted skyrmions in strained chiral magnets, New J. Phys. **20**, 063050 (2018).

[49] C. Psaroudaki and C. Panagopoulos, Skyrmion Qubits: A New Class of Quantum Logic Elements Based on Nanoscale Magnetization, Phys. Rev. Lett. **127**, 067201 (2021).

[50] J. Xia, X. C. Zhang, X. X. Liu, Y. Zhou, and M. Ezawa, Universal Quantum Computation Based on Nanoscale Skyrmion Helicity Qubits in Frustrated Magnets, Phys. Rev. Lett. **130**, 106701 (2023).

[51] K. Wu, Y. L. Zhao, H. Y. Hao, S. Yang, S. Li, Q. F. Liu, S. F. Zhang, X. X. Zhang, J. Akerman, L. Xi, Y. Zhang, K. M. Cai, and Y. Zhou, Topological transformation of synthetic ferromagnetic skyrmions: thermal assisted switching of helicity by spin-orbit torque, Nat. Commun. **15**, 10463 (2024).

[52] X. Yao, J. Chen, and S. Dong, Controlling the helicity of magnetic skyrmions by electrical field in frustrated magnets, New J. Phys. **22**, 083032 (2020).